\documentclass[Physsubmission, Phys]{SciPost}
\hypersetup{
    colorlinks,
    linkcolor={red!50!black},
    citecolor={blue!50!black},
    urlcolor={blue!80!black}
}

\usepackage[bitstream-charter]{mathdesign}
\DeclareSymbolFont{usualmathcal}{OMS}{cmsy}{m}{n}
\DeclareSymbolFontAlphabet{\mathcal}{usualmathcal}

\begin{document}

% TODO: write your article's title here.
% The article title is centered, Large boldface, and should fit in two lines
\begin{center}{\Large \textbf{
Exploring and exploiting various regimes within the jet shower\\
}}\end{center}

% TODO: write the author list here. Use initials + surname format.
% Separate subsequent authors by a comma, omit comma at the end of the list.
% Mark the corresponding author with a superscript *.
\begin{center}
Raghav Kunnawalkam Elayavalli\textsuperscript{1,2} 
\end{center}

% TODO: write all affiliations here.
% Format: institute, city, country
\begin{center}
{\bf 1} Wright Laboratory, Yale University
\\
{\bf 2} Brookhaven National Laboratory
\\
% TODO: provide email address of corresponding author
raghav.ke@yale.edu
\end{center}

\begin{center}
\today
\end{center}

% For convenience during refereeing (optional),
% you can turn on line numbers by uncommenting the next line:
%\linenumbers
% You should run LaTeX twice in order for the line numbers to appear.

\definecolor{palegray}{gray}{0.95}
\begin{center}
\colorbox{palegray}{
  \begin{tabular}{rr}
  \begin{minipage}{0.1\textwidth}
    \includegraphics[width=22mm]{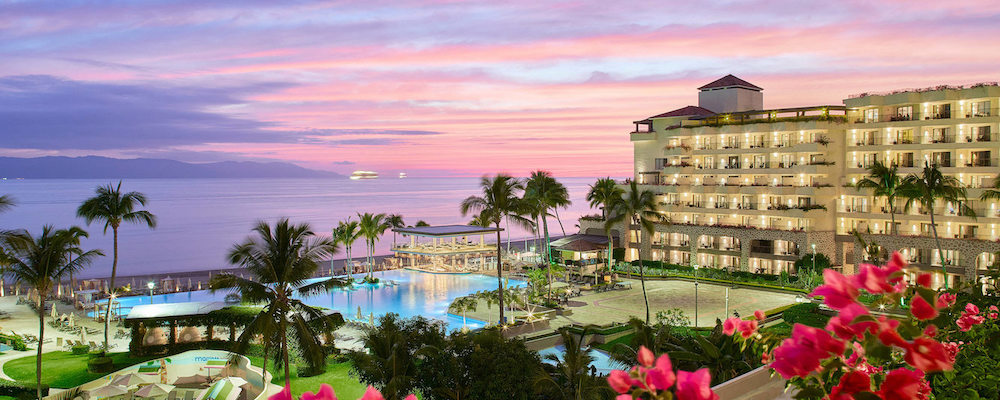}
  \end{minipage}
  &
  \begin{minipage}{0.75\textwidth}
    \begin{center}
    {\it Proceedings for the Winter Workshop on Nuclear Dynamics}\\
    {\it Puerto Vallarta, Feb 2022} \\
    \doi{tbd}\\
    \end{center}
  \end{minipage}
\end{tabular}
}
\end{center}

\section*{Abstract}
{\bf The last few years have seen community-wide excitement in the study of jet substructure derived from the inner workings of clustering algorithms. Such efforts have resulted in the design of new observables which are related to partonic processes from final state hadrons. Since jets are multi-scale objects, they necessarily encode information about both the perturbative (pQCD) parton shower and non-perturbative (npQCD) physics including hadronization. Recent high precision measurements of jet substructure in proton-proton (pp) collisions have pushed the theoretical community into extending their predictions to higher orders resulting in the observation of large theoretical uncertainties from the non-perturbative regime of the calculations. We emphasize the importance of understanding a jet shower from a multi-dimensional point of view and highlight a recent measurement focused on distinguishing the pQCD vs. npQCD regimes within a jet shower. We introduce and discuss the utility of the formation time evaluated at varying stages of the jet shower in pp collisions. Finally, we present a monte-carlo study of the formation time of charged particles within the jet to gain a handle on hadronization mechanisms including string-breaking, and outline a path forward for such observables in heavy ion collisions.  
}

\section{Introduction}
Jets are an algorithmic representation of groups of hadron originating from the fragmentation and hadronization of a hard scattered quark/gluon (parton) in high energy collisions. The latest implementation of jet reconstruction algorithms iteratively clusters objects~\cite{FastJet, Marzani:2019hun}; charged tracks or calorimeter towers in experiment and particles in event generators, within an user-defined angular size commonly referred to as a resolution parameter or radius. These clustering algorithms have proven themselves immensely useful in studying quantum chromodynamics (QCD) at both the Relativistic Heavy Ion Collider (RHIC) and the Large Hadron Collider (LHC). Measured jet cross-sections as a function of the transverse momentum have shown to be adequately reproduced by perturbative calculations (pQCD) across several orders of magnitude~\cite{Britzger:2012bs}. The accuracy of the pQCD predictions at the parton level can be improved with non-perturbative corrections (due to underlying event, multi-parton interactions and hadronization) predominantly estimated from monte-carlo (MC) event generators. Therefore, reconstructed jets are designated as proxies for the out-going hard-scattered partons, and now we study the evolution of the parton via the hadrons associated with the jet. There are two fundamental scales to consider in a parton shower, or simply a splitting process (e.g $q \rightarrow q+g)$, which are the momentum fraction carried by the emitted particle, a gluon in this case, and the emission angle. The splitting probabilities, which are inversely related to the opening angle and momentum fractions, can be converted into the well known DGLAP evolution equations for high energy partons. These evolution equations are utilized in MC to evolve the parton, and once the shower is fully completed and all the remaining partons end up on the mass shell, hadronization mechanisms take over and convert these color charged partons into color neutral hadrons (and decay them as applicable) which can be detected in experiments. The current era of jet measurements are focused on exploiting the information contained within jet clusterings to extract the structure and evolution of the jets.  

There are two main classes of observables related to jet structure measured thus far in high energy collisions; momentum fraction and an opening angle similar to the scales in the parton shower. Canonical jet structure measurements include the jet fragmentation function, i.e. the momentum carried by a hadron (charged particles, for ease of measurement) w.r.t a jet, and the jet shape which measures the (charged) particle density at a certain distance from the jet axis. Both these observables describe the structure of the jet and its fragmentation into hadrons, but do not provide any information related to its evolution. See ~\cite{Cunqueiro:2021wls} for experimental and ~\cite{Blaizot:2015lma} for theoretical review articles. Recently, experimentalists and theorists have come together to expand on the idea of jet clustering and started studying the possibility of translating an ordered collection of particles into a probe of the multi-scale dynamics of the parton shower evolution. One such example of an analysis technique is SoftDrop (SD) grooming~\cite{Larkoski:2014wba, Larkoski:2015lea}, which re-clusters all constituents of a jet into an angular ordered tree via the Cambridge/Aachen (C/A) algorithm. In vacuum, angular ordering follows the emission spectrum wherein subsequent emissions occur at angles smaller than the previous emission. As a consequence, early time emissions can be associated with wide angles and later emission with narrow angles. It has been shown that jet substructure observables at the first hard splitting, such as the momentum fraction ($z_{g}$) and the groomed jet radius ($R_{g}$), can be described sufficiently by pQCD calculations, without the need for large hadronization corrections at sufficiently large jet radii and momenta~\cite{STAR:2020ejj, STAR:2021lvw}. This provides evidence that the early time dynamics of a jet shower can be described in a perturbative fashion while one expects the later stages to be dominated by non-perturbative effects since the end result of a jet shower is hadronized particles measured in the detector. 

\begin{equation}
z_{g} = z = \frac{p_{T, 2}}{p_{T, 1} - p_{T, 2}}; R_{g} = \theta = \Delta R(1, 2)
\label{eq1}
\end{equation}

The definition of the momentum fraction and groomed jet radius as shown in Eq~\ref{eq1} calculated using the transverse momenta ($p_{T}$) and the $y$ (rapidity), $\phi$ (azimuthal angle) of the two objects $(1, 2)$ which are numbered such that $p_{T, 2} < p_{T, 1}$. The opening angle is commonly referred to using the $\theta$ variable, and in this case, is calculated as the distance between the two prongs in the $y-\phi$ space ($\delta R$). For SD, these are the prongs in the C/A clustering that satisfy the grooming criteria, but in principle this definition can be used for any two quantities such as charged particles or other splittings within the clustering tree as we will explore in this study.  

\section{Jet substructure and formation times}

A vacuum parton shower is inherently a quantum-mechanical process and as such the Heisenberg uncertainty principle can be applied at a particular splitting to define a formation time given the energy and angle of the emission. Since the energy scales in discussion are on the order of a few GeVs, we should also include relativistic effects of time dilation and as a consequence, one can define as formation time~\cite{Apolinario:2020uvt} as follows - 

\begin{equation}
\tau_f = \frac{1}{z \cdot (1-z) \cdot \theta^{2} E} [\rm{fm}/c],
\label{eq2}
\end{equation}

where $z$ is the momentum fraction, $\theta$ is the opening angle and $E$ is the energy of the radiator. The impact of such a formation time observable is to associate a time scale to different parts of the jet evolution and to quantify the varying physics regimes in the multi-scale jet shower. We first begin with the perturbative part of the jet shower, i.e the SD first split.  

\begin{figure}
 \includegraphics[width=0.9\linewidth]{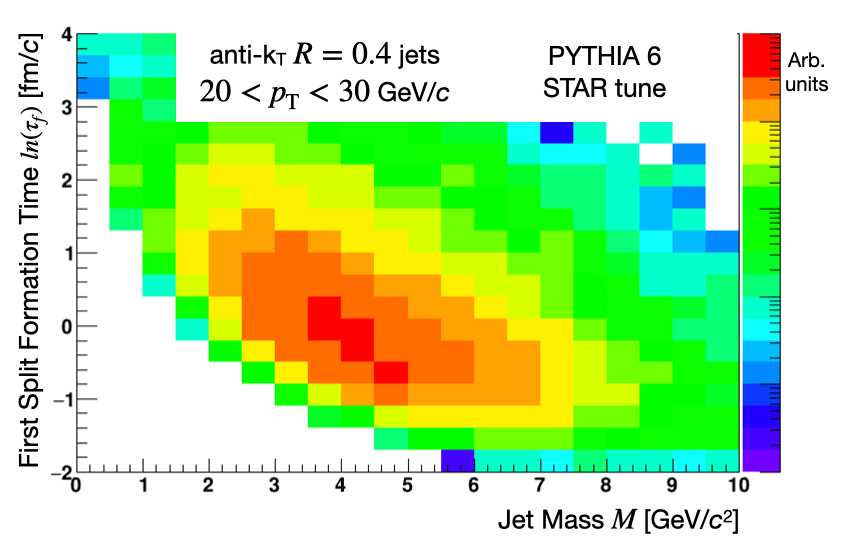}
 \caption{Correlation between the formation time at the first SD emission (y-axis) and the invariant jet mass (x-axis) for jets from PYTHIA 6. The z-axis color scale shows the density of the correlation normalized per total number of jets.}
 \label{fig1}
\end{figure}

At the SD first split, the necessary ingredients to calculate the formation time, such as the momentum fraction ($z = z_{g}$), the opening angle ($\theta = R_{g}$), the mother prong energy ($E$, proxy for jet energy) have been shown in pervious publications~\cite{STAR:2020ejj, STAR:2021lvw}. In order to gain a conceptual understanding of a jet's formation time at the first SD split, we show the correlation between $ln(\tau^{SD}_{f})$ and the invariant jet mass $M = \sqrt{E^{2} - p^{2}}$ in Fig~\ref{fig1}. This correlation is estimated from the PYTHIA 6 event generator~\cite{Sjostrand:2006za} with the STAR tune~\cite{STAR:2018yxi} for anti-$k_{T}~ R=0.4$ jets with $20 < p_{T} < 30$ GeV/$c$. We observe a strong anti-correlation whereby large/late formation times correspond to small jet masses and conversely, those with small/early SD formation times, end up as jets with very large masses. We understand this behavior primarily based on the dependence of both the jet mass and formation time to the opening angle. Early time emissions predominantly occur via a large opening angle and correspond to jets of large virtuality and mass. Jets with a late $\tau^{SD}_{f}$ have a shower topology characterized by very narrow opening angles and subsequently the overall virtuality and mass are smaller on average. 

\begin{figure}
 \includegraphics[width=0.9\linewidth]{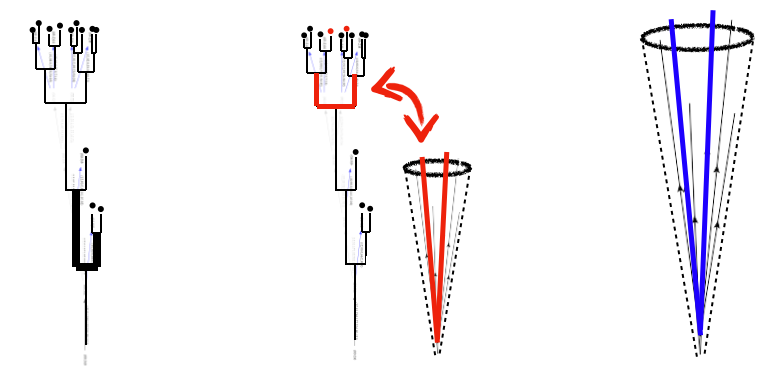}
 \caption{Cartoon representation of the three different stages of calculating the formation time for a particular jet. The left drawing highlights the SD first split in the thick black lines. The right panel shows the leading and subleading charged particles within the jet in the blue solid lines and the middle panel emphasizes the resolved splitting in red, where the charged particles are first resolved by the C/A de-clustering tree into two separate prongs.}
 \label{fig2}
\end{figure}

Now we define, for the first time, three different formation times associated with the jet. The first one is the SD formation time as we discussed earlier, which we expect to follow a pQCD prescription. In order to signify a later segment of the jet shower, we shall consider the leading and sub-leading (in terms of their $p_{T}$) charged particles within the jet as the two objects $1, 2$ and use Eq~\ref{eq2} to calculate a second formation time $\tau^{ch}_f$. This is not a realistic splitting in a theoretical sense, since we know that the charged particles are produced post hadronization. Thus, one can expect that this $\tau^{ch}_f$ will have predominantly later or larger mean values of the formation times as compared to SD first split, and as a consequence, we expect this particular quantity to have large npQCD corrections from hadronization effects~\cite{Chien:2021yol}. Lastly, we look at a third formation time from an intermediate stage within the jet shower which we call the resolved splitting $\tau^{res}_{f}$. The resolved splitting is the first instance along the C/A de-clustering tree wherein the leading and subleading charged particles are separated into their own prongs. This signifies a specific point in time along the jet shower where the high energy hadronized remnants of the jet are resolved in the jet clustering step. All the three types of formation times are shown in Fig~\ref{fig2} where the black markers represent the SD, red for the resolved splittings and blue for the charged particles. 

\begin{figure}
 \includegraphics[width=0.9\linewidth]{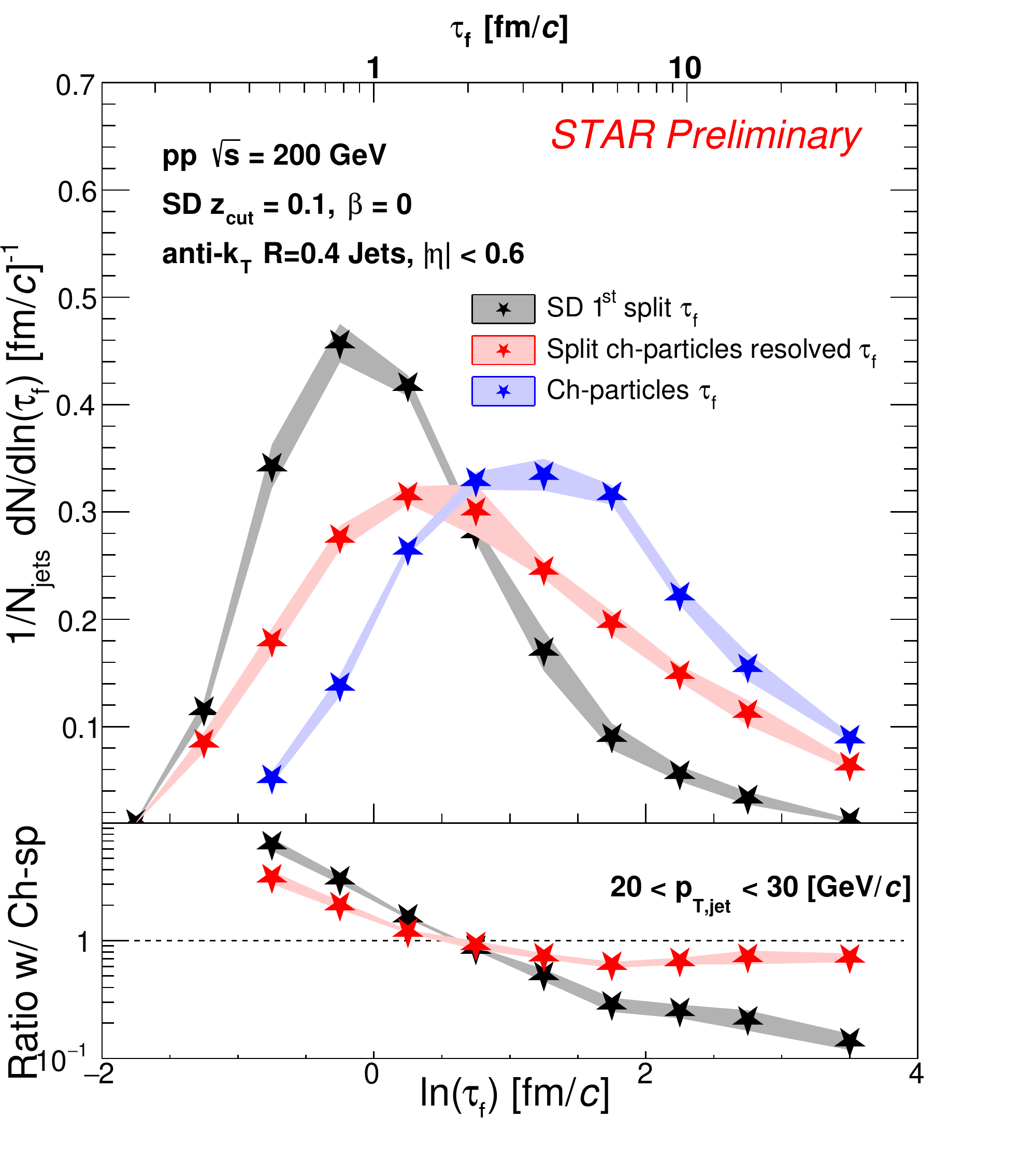}
 \caption{Fully corrected preliminary measurement from the STAR collaboration of the formation time for the SD first splitting (black), resolved splitting (red) and the leading and subleading charged particle (blue). The shaded regions behind the data markers represent the total systematic uncertainties. The formation times are calculated for anti$-k_{T}~ R=0.4$ jets with $20 < p_{T} < 30$ GeV/$c$ and the bottom panel shows the ratio of SD and resolved splittings to the charged-particle split. %Past a formation time of $ \tau_f \approx 3-4$ fm, we observe the ratio to be flat between the resolved splittings and the ch-particles.
 }
 \label{fig3}
\end{figure}

The first measurement of these formation time observables within a jet shower was done by the STAR collaboration at RHIC and presented recently for anti$-k_{T}~ R=0.4$ jets with $20 < p_{T} < 30$ GeV/$c$ in proton-proton collisions at $\sqrt{s} = 200$ GeV. The top panel shows the fully corrected distributions of the $\tau^{SD}_{f}$ in black markers, $\tau^{res}_{f}$ in red markers and $\tau^{ch}_{f}$ in blue markers, where each distribution is normalized to its integral. As expected, we observe a trend in the three different formation times with $\langle \tau^{SD}_{f} \rangle < \langle \tau^{res}_{f} \rangle < \langle \tau^{ch}_{f} \rangle$. In order to quantify the shapes of these formation times, the bottom panel in Fig~\ref{fig3} presents the ratio of both the $\tau^{SD}_{f}$ and $\tau^{res}_{f}$ with $\tau^{ch}_{f}$. At early formation times, we see a large negative slope for the $\tau^{SD}_{f}$ splittings indicating the early time (perturbative) peak of the SD selected splits. At later formation times ($\tau > 3 - 4 $ fm/$c$), the shape of the resolved splittings are nearly consistent with the charged hadron splittings. With such specially selected resolved splitting, we can now identify and specify a time within the jet clustering shower where all the splittings that come afterwards, can be described as the splitting one calculated using hadrons as opposed to a partonic splitting. This measurement now provides a unique method whereby one can use the resolved splittings within a jet shower, to select a specific jet topology entirely based on the time a jet spends in the perturbative sector ($\tau^{res}_{f} < 2$ fm/$c$), or in the non-perturbative regime ($\tau^{res}_{f} > 5$ fm/$c$). These results will serve as a baseline in proton-proton collisions to first identify the region of perturbative calculability and then compare different hadronization mechanisms to explore the non-perturbative sector in QCD. 

Space-time tomography of the quark-gluon plasma (QGP) via jets produced in heavy ion collisions is one of the standard methods of studying the transport properties of deconfined quarks and gluons. Since the jets evolve concurrently with the QGP, selecting jet topologies with a particular formation time enables a time-dependent study of the jet-QGP interactions. Since there is a large heavy ion underlying event there is a requirement that the selection of the formation time observable in a jet be robust to the background and still be sensitive to the kinematics of the jet shower. The first SD split for jets in heavy ion collisions, with the nominal value of the grooming parameters, have been shown to be extremely sensitive to the heavy ion background~\cite{STAR:2021kjt}. This can be overcome by utilizing a stricter grooming criterion~\cite{Mulligan:2020tim} which affects the selection bias of the surviving jet population. Since the leading and subleading charged particles in a jet are predominantly higher momentum, their reconstruction is unaffected by the presence of the underlying event.  Thus, the resolved splitting is expected to be a robust guide to selecting specific jet topologies and a formation time which can be translated to their path-length traversed in the QGP medium. 

\section{Hadronization and jet substructure}

\begin{figure}
 \includegraphics[width=0.49\linewidth]{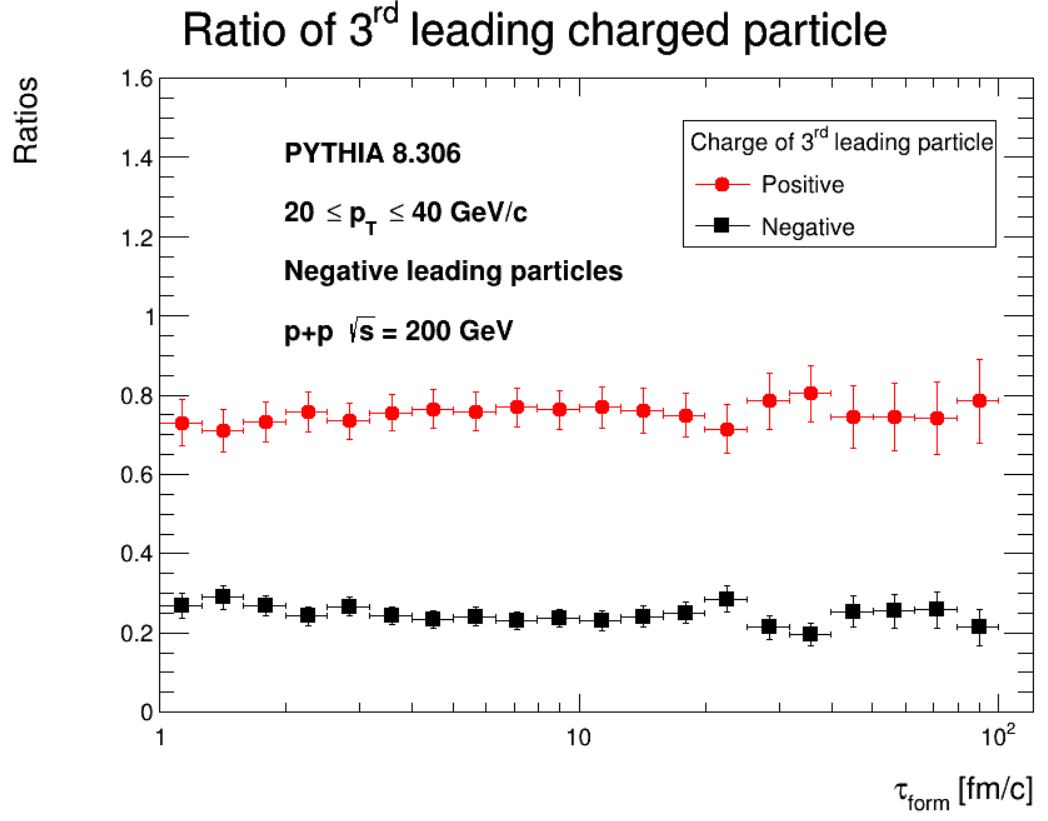}
 \includegraphics[width=0.49\linewidth]{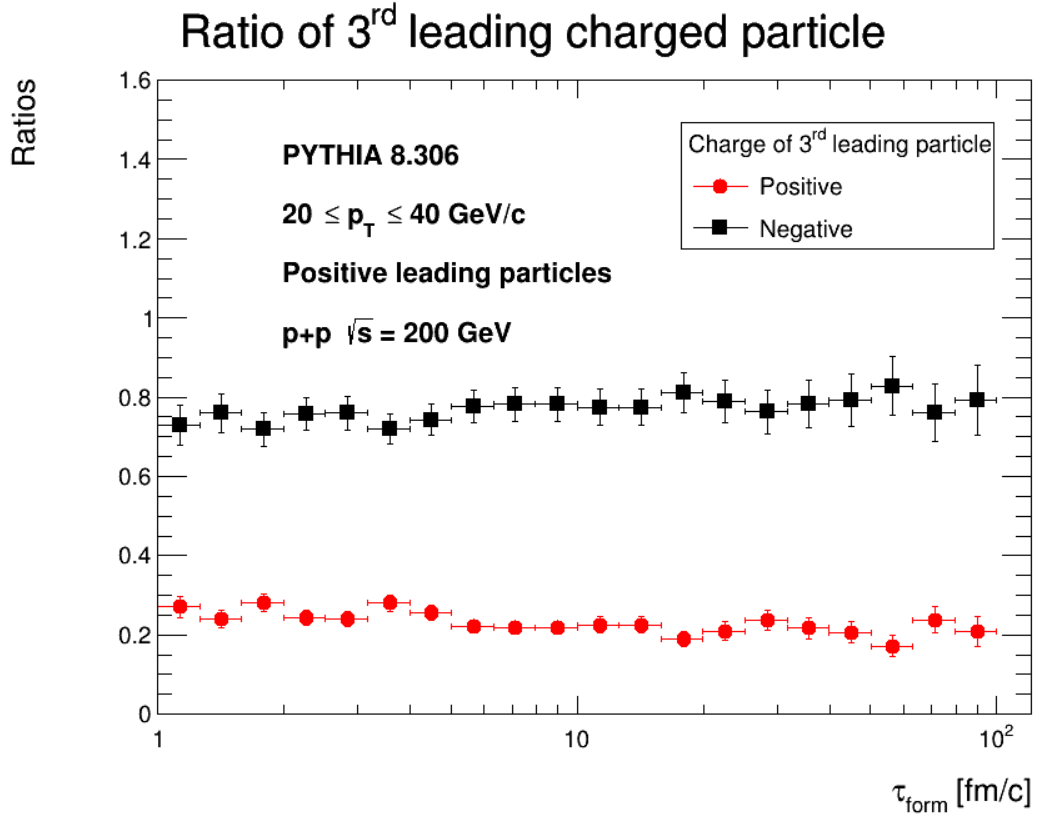}
 \caption{Top: Normalized ratio of ch-particle formation time when the leading and subleading charged particles are negatively charged and the third particle, ordered in $p_{T}$ is either positive (red circles) or negative (black boxes) as a function of formation time. Bottom: Similar ratio of ch-particle formation time in the scenario that the leading and subleading charged particles are positively charged and the third highest $p_{T}$ particle is either positive (red) or negative (black).}
 \label{fig4}
\end{figure}

In order to study the impact of selecting on the charged hadron formation time, we look at the charge of the particles which make up the first, second and third (ordered in $p_{T}$) charged hadrons in the jet constituents. The charged particle formation time is then plotted for same charged leading and subleading hadrons with the third particle being either positively or negatively charged. The ratios of these formation times are shown in Fig~\ref{fig4} for negatively (top) and positively (bottom) charged leading and subleading charged particles. The red circle (black box) markers in each of the panels correspond to the third charged particle being positively (negatively) charged. In each of the cases, we observe a clear separation of the formation times based on the charge of the third particle predominantly expected to oppositely charged to the leading and subleading particles. There is no dependence to the formation time on this ratio estimated from PYTHIA 8 simulations~\cite{Bierlich:2022pfr} at jet kinematics accessible at RHIC energies. This particular separation shown in Fig~\ref{fig4} is a feature of the Lund string breaking hadronization mechanism implemented in PYTHIA where one expects an overall balancing of electric charge in the hadrons produced due to the jet fragmentation. Ongoing experimental measurements at both RHIC and LHC are expected to highlight this feature and aim to understand hadronization mechanism via data-driven approaches. 

\section{Conclusions}

We defined the concept of jet substructure and introduced many of the recently studied analysis techniques such as SoftDrop and iterative clustering. We also introduced the idea of formation times at varying stages of the jet shower and presented the selection of the resolved splittings. Recent STAR data of the formation times corresponding to the SD first splits, leading and subleading charged hadron and the resolved splits was discussed. The resolved splitting is shown to identify a particular time within the jet clustering tree wherein the splittings are consistent with the formation time estimated via hadronized charged particles. For jets at RHIC energies, the transition region from pQCD to npQCD is expected to occur $\tau \approx 3-5$ fm/$c$. We finally highlighted the ability of such formation time observables to potentially study hadronization mechanisms in data by looking at the dependence of the electric charge of the particles considered in the calculation of the charged particle formation time. This work was supported by the Office of Nuclear Physics of the U.S. Department of Energy under award number DE-SC004168.

\nolinenumbers


\begin{thebibliography}{99}

\bibitem{FastJet} M. Cacciari, G. P. Salam, and G. Soyez, J. High Energy Phys. 04 (2008) 005.

\bibitem{Marzani:2019hun}
S.~Marzani, G.~Soyez and M.~Spannowsky,
%``Looking inside jets: an introduction to jet substructure and boosted-object phenomenology,''
Lect. Notes Phys. \textbf{958}, pp. (2019)
%doi:10.1007/978-3-030-15709-8
%[arXiv:1901.10342 [hep-ph]].
%66 citations counted in INSPIRE as of 19 Jul 2021

\bibitem{Britzger:2012bs}
D.~Britzger \textit{et al.} [fastNLO],
%``New features in version 2 of the fastNLO project,''
%doi:10.3204/DESY-PROC-2012-02/165
arXiv:1208.3641
%134 citations counted in INSPIRE as of 28 Jun 2022

\bibitem{Cunqueiro:2021wls}
L.~Cunqueiro and A.~M.~Sickles,
%``Studying the QGP with Jets at the LHC and RHIC,''
Prog. Part. Nucl. Phys. \textbf{124}, 103940 (2022)
%doi:10.1016/j.ppnp.2022.103940
%[arXiv:2110.14490 [nucl-ex]].
%15 citations counted in INSPIRE as of 28 Jun 2022

%\cite{Blaizot:2015lma}
\bibitem{Blaizot:2015lma}
J.~P.~Blaizot and Y.~Mehtar-Tani,
%``Jet Structure in Heavy Ion Collisions,''
Int. J. Mod. Phys. E \textbf{24}, no.11, 1530012 (2015)
%122 citations counted in INSPIRE as of 28 Jun 2022

%\cite{Larkoski:2014wba}
\bibitem{Larkoski:2014wba}
A.~J.~Larkoski, S.~Marzani, G.~Soyez and J.~Thaler,
%``Soft Drop,''
JHEP \textbf{05}, 146 (2014)
%doi:10.1007/JHEP05(2014)146
%[arXiv:1402.2657 [hep-ph]].
%622 citations counted in INSPIRE as of 19 Jul 2021


%\cite{Larkoski:2015lea}
\bibitem{Larkoski:2015lea}
A.~J.~Larkoski, S.~Marzani and J.~Thaler,
%``Sudakov Safety in Perturbative QCD,''
Phys. Rev. D \textbf{91}, no.11, 111501 (2015)
%doi:10.1103/PhysRevD.91.111501
%[arXiv:1502.01719 [hep-ph]].
%104 citations counted in INSPIRE as of 19 Jul 2021

%\cite{STAR:2020ejj}
\bibitem{STAR:2020ejj}
J.~Adam \textit{et al.} [STAR],
%``Measurement of groomed jet substructure observables in p+p collisions at $\sqrt {s}$ =200 GeV with STAR,''
Phys. Lett. B \textbf{811}, 135846 (2020)
%doi:10.1016/j.physletb.2020.135846
%[arXiv:2003.02114 [hep-ex]].
%17 citations counted in INSPIRE as of 19 Jul 2021



%\cite{STAR:2021lvw}
\bibitem{STAR:2021lvw}
M.~Abdallah \textit{et al.} [STAR],
%``Invariant Jet Mass Measurements in $pp$ Collisions at $\sqrt{s} = 200$ GeV at RHIC,''
Phys. Rev. D \textbf{104}, no.5, 052007 (2021)
%5 citations counted in INSPIRE as of 28 Jun 2022

%\cite{Apolinario:2020uvt}
\bibitem{Apolinario:2020uvt}
L.~Apolin\'ario, A.~Cordeiro and K.~Zapp,
%``Time reclustering for jet quenching studies,''
Eur. Phys. J. C \textbf{81}, no.6, 561 (2021)
%9 citations counted in INSPIRE as of 28 Jun 2022

%\cite{Sjostrand:2006za}
\bibitem{Sjostrand:2006za}
T.~Sjostrand, S.~Mrenna and P.~Z.~Skands,
%``PYTHIA 6.4 Physics and Manual,''
JHEP \textbf{05}, 026 (2006)
%11912 citations counted in INSPIRE as of 20 Jul 2021


%\cite{STAR:2018yxi}
\bibitem{STAR:2018yxi}
J.~Adam \textit{et al.} [STAR],
%``Longitudinal double-spin asymmetries for dijet production at intermediate pseudorapidity in polarized $pp$ collisions at $\sqrt{s}=$ 200 GeV,''
Phys. Rev. D \textbf{98}, no.3, 032011 (2018)
%19 citations counted in INSPIRE as of 20 Jul 2021

%\cite{Chien:2021yol}
\bibitem{Chien:2021yol}
Y.~T.~Chien, A.~Deshpande, M.~M.~Mondal and G.~Sterman,
%``Probing hadronization with flavor correlations of leading particles in jets,''
Phys. Rev. D \textbf{105}, no.5, L051502 (2022)
%3 citations counted in INSPIRE as of 28 Jun 2022


%\cite{STAR:2021kjt}
\bibitem{STAR:2021kjt}
M.~S.~Abdallah \textit{et al.} [STAR],
%``Differential measurements of jet substructure and partonic energy loss in Au+Au collisions at $\sqrt {S_{NN}}$ =200 GeV,''
Phys. Rev. C \textbf{105}, no.4, 044906 (2022)
%7 citations counted in INSPIRE as of 28 Jun 2022


\bibitem{Mulligan:2020tim}
J.~Mulligan and M.~Ploskon,
%``Identifying groomed jet splittings in heavy-ion collisions,''
Phys. Rev. C \textbf{102}, no.4, 044913 (2020)
%20 citations counted in INSPIRE as of 28 Jun 2022

\bibitem{Bierlich:2022pfr}
C.~Bierlich, S.~Chakraborty, N.~Desai, L.~Gellersen, I.~Helenius, P.~Ilten, L.~L\"onnblad, S.~Mrenna, S.~Prestel and C.~T.~Preuss, \textit{et al.}
%``A comprehensive guide to the physics and usage of PYTHIA 8.3,''
arXiv:2203.11601
%9 citations counted in INSPIRE as of 28 Jun 2022


\end{thebibliography}
\end{document}